\documentclass[prl,aps,twocolumn,showpacs,superscriptaddress,floatfix]{revtex4-2}

\usepackage{mathtools}
\usepackage[usenames,dvipsnames]{xcolor}
\usepackage{amsmath}
\usepackage{amssymb,mathrsfs}
\usepackage{graphicx}
\usepackage[colorlinks=true]{hyperref}
\usepackage{soul}
\usepackage{xcolor}

\usepackage{float}

\newcommand{\prt}{\partial}

\newcommand{\eps}{\varepsilon}

\newcommand{\om}{\omega}

\newcommand{\vphi}{\varphi}

\newcommand{\sn}{\mathrm{sn}}

\newcommand{\sgn}{\mathrm{sgn}}

\begin{document}

\title{`Relativistic' propagation of instability fronts in nonlinear Klein-Gordon equation dynamics}

\author{A. M. Kamchatnov}
\affiliation{Institute of Spectroscopy, Russian Academy of Sciences, Moscow, Troitsk, 108840, Russia}
%\affiliation{ Higher School of Economics, Physical Department, 20 Myasnitskaya ulica, Moscow 101000, Russia}

\begin{abstract}
We consider propagation of instability fronts in conservative nonlinear wave systems by the
Whitham method. Whitham modulation equations for periodic solutions of the generalized Klein-Gordon
equation are solved in the limit of asymptotically large times, when the size of the instability wave
region is much greater than the size of the initial localized disturbance, so the solution reaches
the self-similar regime. The general self-similar solution is illustrated by two typical examples
of the nonlinearity function. It is shown that in these models the instability fronts propagate 
with maximal group velocity.
\end{abstract}


% 05.45.−a Nonlinear dynamics and chaos
% 05.45.Yv Solitons
% 47.35.Fg Solitary waves

\maketitle

%\section{Introduction}

{\it 1. Introduction.}---Propagation of instability fronts is a universal phenomenon manifesting 
itself in various physical situations (see, e.g., review article \cite{vanSaar-03} and references 
therein). In a number of typical situations, a localized perturbation is initially confined to a 
small region, and then it spreads out into the rest part of an unstable system. This spreading 
often occurs with a well-defined velocity defined by the intrinsic properties of the system, 
independent of the initial conditions. In particular, in the case of nonlinear diffusion systems, 
these velocities correspond to the `marginal stability' condition \cite{dl-83,vanSaar-87} similar 
to the condition that separates the absolute instability from the convective one \cite{sturrock,briggs}. 
The same selection rule for the instability propagation velocity remains correct in case of 
non-dissipative modulationally unstable nonlinear dispersive systems. For example, 
a plane wave, described by a nonlinear Schr\"{o}dinger (NLS) equation, is modulationally unstable 
\cite{ostrovsky-66,bf-67,zakharov-68} with respect to disintegration into wave packets. Therefore, 
a small localized perturbation of a plane wave leads to the formation of a 
region of nonlinear oscillations whose small-amplitude edges propagate into the undisturbed part of 
the wave with the minimal group velocity of the propagating mode \cite{kk-69,karpman} in agreement 
with the `marginal stability' condition. This observation was confirmed by the solution of the 
Whitham modulation equations for the whole region of oscillations \cite{kamch-92,egkk-93,bk-94,kamch-97}. 
A similar situation takes place for the stability of dark solitons described by the two-dimensional 
NLS equation \cite{kt-88}, so that the transition from absolute instability to convective one 
explains the existence of oblique solitons in a flow of a Bose-Einstein condensate past an 
obstacle \cite{egk-06,kp-08,kk-11} in agreement with experimental observations in 
Refs.~\cite{amo-11,grosso-11}. In this respect, it is especially interesting that the property of 
`marginal stability' is captured by the exact self-similar solutions of the Whitham modulation 
equations \cite{whitham-65,whitham} derived for periodic solutions of the NLS equation in 
Refs.~\cite{fl-86,pavlov-87}. Actually, the solutions of Refs.~\cite{kamch-92,egkk-93,bk-94} 
(see also Ref.~\cite{kamch-97}) provide examples of the exact realization of the general considerations 
presented in Refs.~\cite{dl-83,vanSaar-87} specified for some class of completely integrable equations. 
This observation suggests that the Whitham equations can be applied to the description of the propagation 
of instability fronts of other systems. In this Letter, we will apply this method to the generalized 
nonlinear Klein-Gordon equation and find a remarkably simple self-similar solution of Whitham equations 
that describes the propagation of instability fronts. According to this solution, the instability
fronts propagate with maximal group velocity, corresponding to the short wavelength limit of the
dispersion relation.

{\it 2. Periodic solutions and modulation equations.}---A paradigmatic example of a modulationally
unstable situation is given by the well-known nonlinear Klein-Gordon equation,
\begin{equation}\label{eq1}
  \vphi_{tt}-\vphi_{xx}+U'(\vphi)=0,\quad U'=\frac{dU}{d\vphi},\quad U(0)=0,
\end{equation}
with a nonlinear `potential' $U(\vphi)$.  
It is easy to see that Eq.~(\ref{eq1}) has travelling wave solutions $\vphi=\vphi(\xi)$, $\xi=x-Vt$,
where $\vphi(\xi)$ is defined implicitly by the formula
\begin{equation}\label{eq2}
  \xi=\sqrt{\frac{V^2-1}{2}}\int_{\vphi_0}^{\vphi}\frac{d\vphi}{\sqrt{A-U(\vphi)}},
\end{equation}
with $\vphi(0,0)=\vphi_0$. The variable $\vphi$ oscillates between the turning points given by
the two roots of the equation $A-U(\vphi)=0$, so it can be called the amplitude parameter. 
In a modulated wave, the amplitude parameter $A$ and the
phase velocity $V$ become slow functions of $x$ and $t$, and their dynamics is determined by the modulation
equations derived by Whitham in Ref.~\cite{whitham-65}. According to Whitham, it is convenient to express
the modulation equations in terms of the function
\begin{equation}\label{eq3}
\begin{split}
 & W(V,A)=\sqrt{V^2-1}\,G(A),\\
 & G(A)=\sqrt{2}\oint \sqrt{A-U(\vphi)}\,d\vphi,
  \end{split}
\end{equation}
where the integration is taken over the interval $\vphi\in\{\vphi: A-U(\vphi)\geq0\}$. Then the wavelength
is equal to
\begin{equation}\label{eq4}
  L=\frac{\prt W}{\prt A}=\sqrt{V^2-1}\cdot G'(A),
\end{equation}
so the wave number $k=1/L$ and the frequency $\om=kV$ are related with each other by the dispersion relation
\begin{equation}\label{eq5}
  \om^2=k^2+(G'(A))^{-2}.
\end{equation}
Consequently, the group velocity is equal to
\begin{equation}\label{eq6}
  v=\left(\frac{\prt\om}{\prt k}\right)_A=\frac{k}{\om}=\frac{1}{V}.
\end{equation}
As is clear from Eq.~(\ref{eq3}), we always have $V>1$, so the physical velocity $v$ of the signal's propagation
is always smaller than unity.

Whitham derived the modulation equations in the form
\begin{equation}\label{eq7}
  \begin{split}
  & \left(\frac{kV}{V^2-1}+A\right)_t+\left(\frac{kVW}{V^2-1}\right)_x=0,\\
  & \left(\frac{kVW}{V^2-1}\right)_t+\left(\frac{kV^2W}{V^2-1}-A\right)_x=0.
  \end{split}
\end{equation}
It is convenient to exclude the variables $W$ and $V$ with the help of Eqs.~(\ref{eq3}) and (\ref{eq6}), so we get
\begin{equation}\label{eq8}
  \begin{split}
  & \left(\frac{G/G'}{1-v^2}+A-\frac{G}{G'}\right)_t+\left(\frac{(G/G')v}{1-v^2}\right)_x=0,\\
  & \left(\frac{(G/G')v}{1-v^2}\right)_t+\left(\frac{(G/G')v^2}{1-v^2}-A+\frac{G}{G'}\right)_x=0.
  \end{split}
\end{equation}
Linearization of these equations with respect to small deviations of $A$ and $v$ from constant `background'
values yields the characteristic velocities
\begin{equation}\label{eq9}
  v_{\pm}=\frac{v\pm c}{1\pm vc},
\end{equation}
where $c$ is defined by the expression
\begin{equation}\label{eq10}
  c^2=-{GG^{\prime\prime}}/{(G')^2}.
\end{equation}
If $c$ is real, then Eqs.~(\ref{eq9}) are obviously interpreted as velocities of upstream and downstream sound
propagation with the relativistic rule of addition of velocities. In this case, Eqs.~(\ref{eq8}) can be rewritten
in the form of equations of relativistic hydrodynamics (see, e.g., Ref.~\cite{kamch-23}). If $c^2<0$ and the 
characteristic velocities are complex, then such a wave is modulationally unstable, so its smooth perturbation 
grows with time. This is just the situation we are interested in.

{\it 3. Evolution of a small localized perturbation.}---Let us assume that the potential $U(\vphi$) has a 
minimum at $\vphi=0$, and it grows monotonously with $\vphi$ up to some maximal value at $\vphi=\vphi_m$. 
Then the state $\vphi=0$ is stable and the state $\vphi=\vphi_m$ is obviously unstable. If we disturb this 
unstable state locally in the vicinity of the point $x=0$, then the left- and right-propagating waves 
%along negative and positive directions of the $x$-axis 
will form an oscillatory structure between negative 
and positive instability fronts. We assume that this oscillatory structure can be represented as a modulated 
nonlinear periodic solution of Eq.~(\ref{eq1}), so the solution of the Whitham modulation Eqs.~(\ref{eq8}) 
yields the profiles of the amplitude parameter $A$ and the group velocity $v$ in the whole region, including 
the edges at the instability fronts.

The assumption that the initial disturbance is small suggests that its size becomes inessential very fast,
so the solution of Eqs.~(\ref{eq8}) becomes self-similar, and from dimensional considerations we conclude
that the flow velocity distribution is given by the formula
\begin{equation}\label{eq11}
  v=x/t.
\end{equation}
Besides that, it is natural to assume that the amplitude parameter $A=A(x,t)$ only depends on the
`relativistic' interval: $A=A(z)$, $z=t^2-x^2$. Making these two assumptions, we reduce each of 
Eqs.~(\ref{eq8}) to the same ordinary differential equation 
$G(A)/(zG'(A))+2dA/dz=0$,
%\begin{equation}\label{eq12}
%  \frac{1}{z}\frac{G(A)}{G'(A)}+2\frac{dA}{dz}=0
%\end{equation}
which is readily solved to give
\begin{equation}\label{eq13}
  G(A)={C}(t^2-x^2)^{-1/2},
\end{equation}
where $C$ is an integration constant. This formula defines implicitly the dependence of the amplitude 
$A$ on $x$ and $t$. The applicability condition of the Whitham modulation theory for a concrete function 
$G(A)$ demands that the integration constant should take such values that the relevant intervals of $x$ 
contain a large number of oscillations. To put this in another way, we can choose $C$ equal to unity, 
assuming that $x$ is replaced by $x/\eps$, so that the wavelength (\ref{eq4}) acquires a small factor 
$\eps\ll1$. This means that the Whitham equations (\ref{eq7}) or (\ref{eq8}) are written for a large-scale 
evolution of envelope functions with characteristic size about unity. We will choose this definition of 
scales and take $C=1$ in what follows.

Now, we will illustrate the solution (\ref{eq11}), (\ref{eq13}) by concrete examples.

%\smallskip

\begin{figure}[t]
    \centering
    \includegraphics[width=8cm]{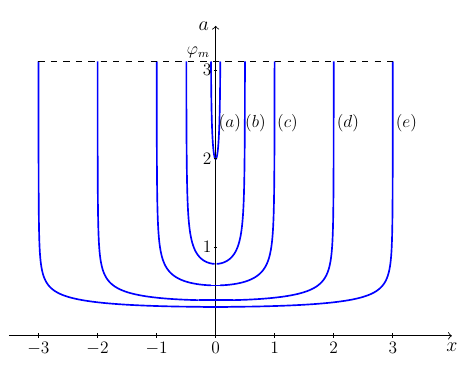}
    \caption{Envelopes of amplitudes $a$ for the sine-Gordon instability wave at different moments
    of time: (a) $t=0.1$; (b) $t=0.5$; (c) $t=1$; (d) $t=2$; (e) $t=3$. The dashed line corresponds to the
    maximal amplitude $\vphi_m=\pi$.}
    \label{fig1}
\end{figure}

\underline{(a) sine-Gordon equation.} If $U(\vphi)=1-\cos\vphi$, then the integral in Eq.~(\ref{eq3})
is readily calculated, % (see, e.g., Ref.~\cite{kamch-23}),
\begin{equation}\label{eq14}
  G(A)=16\left[E(A/2)-(1-A/2)K(A/2)\right],
\end{equation}
where $K(m)$ and $E(m)$ are the complete elliptic integrals of the first and second kind, respectively.
The squared `sound velocity' (see Eq.~(\ref{eq10}))
\begin{equation}\label{eq15}
  c^2=-4A(2-A)\left(\frac{1}{K(A/2)}\frac{dK(A/2)}{dA}\right)^2<0
\end{equation}
is negative for all physically acceptable values of $A$, $0\leq A\leq2$, that correspond to 
$-\vphi_m\leq \vphi\leq \vphi_m$, $\vphi_m=\pi$. The integral in Eq.~(\ref{eq3}) reduces to the 
elliptic one, and its inversion yields the periodic solution in the explicit form
\begin{equation}\label{eq16}
\begin{split}
  \vphi  =2\arcsin\left[\sqrt{\frac{A}2}\,\,\sn\left(\frac{vx-t}{\sqrt{1-v^2}},\frac{A}2\right)\right].
  \end{split}
\end{equation}
Consequently, small values of $A$ correspond to small-amplitude waves around the stable state $\vphi=0$.
In the limit $A\to2$, this solution converts into the kink or anti-kink solution 
of the sine-Gordon equation,
\begin{equation}\label{eq17}
  \vphi=2\arcsin\left[\pm\tanh\left(\frac{vx-t}{\sqrt{1-v^2}}\right)\right],
\end{equation}
so that $\vphi=-\pi$ as $x\to-\infty$ and $\vphi=\pi$ as $x\to+\infty$ for the upper sign in Eq.~(\ref{eq17})
and vice versa for the lower sign. Consequently, this solution connects two unstable states at
the maxima of the potential $U(\vphi)=1-\cos\vphi$, and for smaller values $A<2$ we have nonlinear oscillations
with the amplitude 
\begin{equation}\label{eq18}
  a=2\arcsin\sqrt{A/2}.
\end{equation}
Combining it with the formula
\begin{equation}\label{eq19}
  (t^2-x^2)^{-1/2}=16\left[E(A/2)-(1-A/2)K(A/2)\right],
\end{equation}
we obtain the dependence of $a$ on $x$ for any fixed moment of time $t$ in a parametric form. 
The corresponding plots are shown in Fig.~\ref{fig1}. As one can see, the instability region expands 
with time, and its fronts at the edges move (in the Whitham theory approximation) with the `light' 
velocity $v=1$. In the vicinity of the edge points the wave consists of `trains' of kinks with 
the amplitudes close to the maximal value $\vphi_m=\pi$.

\begin{figure}[t]
    \centering
    \includegraphics[width=8cm]{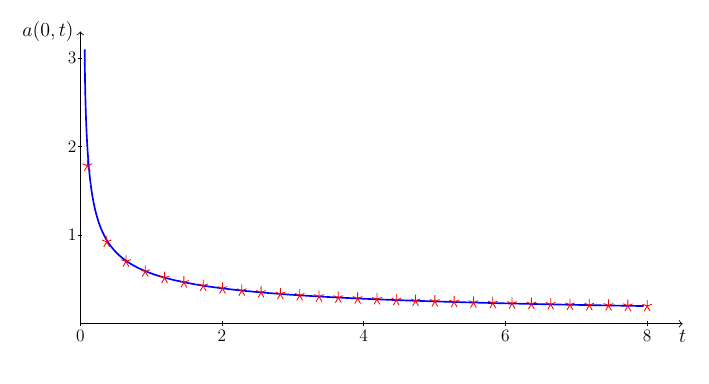}
    \caption{Dependence of the amplitude $a(0,t)$ at the center of the expanding wave structure on time $t$.
    Stars correspond to the asymptotic formula (\ref{eq20}), so it provides a good enough approximation 
    for $t\gtrsim0.5$.}
    \label{fig2}
\end{figure}

The amplitude decreases at the center of the wave structure which becomes here a wave of oscillations around
the stable state $\vphi=0$. This decrease of the amplitude at the center is quite slow (see Fig.~\ref{fig2})
and for large $t$ (small $A\ll1$) Eqs.~(\ref{eq18}) and (\ref{eq19}) give
\begin{equation}\label{eq20}
  a\approx\frac{1}{\sqrt{\pi t}}.
\end{equation}
However, this asymptotic formula provides a very good approximation even for relatively large amplitudes $a$
(the stars in Fig.~\ref{fig2} correspond to this formula).
It is worth noticing that since $G(A)\leq16$, Eq.~(\ref{eq19}) is only correct for
$(t^2-x^2)^{-1/2}<16$, so we must have $t>1/16$ for points at the center of the distribution---the Whitham method is
asymptotical by definition, and its formal continuation to the scales much smaller than unity can go beyond
its applicability region, as it happens here for the limit $t\to0$.

%\smallskip

\begin{figure}[t]
    \centering
    \includegraphics[width=8cm]{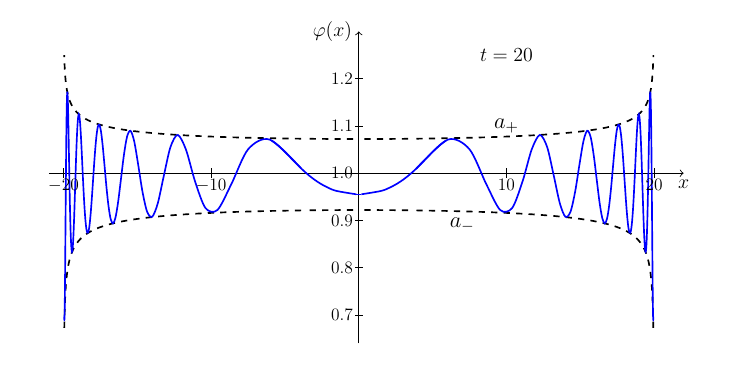}
    \caption{Profile of the instability waves at $t=20$ in the two-well potential model. 
    The modulated wave (\ref{eq26}) is
    depicted by a blue line and the enveloped $a_{\pm}$ following from Eq.~(\ref{eq27}) are shown by
    dashed lines.}
    \label{fig3}
\end{figure}

\underline{(b) Two-well potential model.} Now we assume that the potential $U(\vphi)$ is given by a 
fourth-degree polynomial of $\vphi$. Let initially $U(\vphi)$ have a single minimum at $\vphi=0$, 
which corresponds to a stable state of the system. Then the coefficient before the quadratic term 
changes its sign, so this state becomes unstable, and we have two other stable states with non-zero 
values of $\vphi$. For simplicity, we assume that the potential acquires the form
\begin{equation}\label{eq24}
  U(\vphi)= (\vphi^2-1)^2,
\end{equation}
so the two new `vacua' correspond to $\vphi=\pm1$. If we disturb the unstable state $\vphi=0$ in the vicinity
of the point $x=0$, then two waves of instability start to propagate away, and we assume again that they can be
represented as a modulated periodic solution of Eq.~(\ref{eq1}). In case of transition to the vacuum $\vphi=1$,
the periodic solution corresponds to oscillations of $\vphi$ in the interval
\begin{equation}\label{eq25}
  a_-=\sqrt{a_+^2-1}\leq\vphi\leq a_+, \quad 1<a_+<\sqrt{2},
\end{equation}
where $a_-$ and $a_+$ are the lower and upper envelopes of the wave amplitude. It is easy to express the 
periodic solution in terms of the Jacobi elliptic $\sn$-function,
\begin{equation}\label{eq26}
  \vphi=\left\{a_+^2-2(a_+^2-1)\sn^2\left[\frac{a_+(vx-t)}{\sqrt{1-v^2}},m\right]\right\}^{1/2},
\end{equation}
where $m=2(1-1/a_+^2)$.
In a modulated wave, $v$ and $a_+$ become functions of $x$ and $t$ which change little along one wavelength.
In our self-similar solution of the Whitham modulation equations we have $v=x/t$, and dependence of $a_+$
on $t^2-x^2$ is defined by the equation $G(a_+)=(t^2-x^2)^{-1/2}$ (see Eq.~(\ref{eq13})), where
\begin{equation}\label{eq27}
  G(a_+)=\frac{4\sqrt{2}}{3}a_+\left\{E(m)- (2-a_+^2)K(m)\right\}.
\end{equation}
Plots of the modulated waves of instability and of envelopes of the oscillation amplitudes at some moment 
of time are shown in Fig.~\ref{fig3}. The instability fronts propagate with the maximal group velocities 
$v=\pm1$ velocity, and in between we have oscillations around the new vacuum $\vphi=1$.

{\it 4. Discussion.} It is well known that the Whitham modulation theory provides quite a general 
approach to studying the stability of nonlinear waves (see, e.g., \cite{whitham,ir-2000}). In the case
of stable waves, the Whitham method forms the basis for the theory of dispersive shock waves \cite{gp-73}
(see also Refs.~\cite{eh-16,kamch-21}). It was indicated long ago \cite{lighthill-65} that this method
is also applicable to modulationally unstable systems, and some concrete examples were obtained for the
completely integrable focusing NLS \cite{kamch-92,egkk-93,bk-94,kamch-97} and sine-Gordon
\cite{kk-76,fml-83,ks-91,novok-96,kamch-23} equations. Usually, the solutions considered in these
papers corresponded to different boundary conditions at $x\to\pm\infty$. In this Letter, we discussed
a problem of evolution of an initially localized disturbance in a modulationally unstable system,
and we showed for the case of the generalized Klein-Gordon equation that such a problem admits a simple
self-similar solution. We showed that this solution of the Whitham equations yields the complete 
description of the instability waves propagating into the unstable regions. The edges of such 
waves correspond to the instability fronts, which propagate with maximal group velocity $v\to1$
reached in the short wavelength limit $k\to\infty$.

\end{document}